\documentclass{article}

\usepackage{cite}
\usepackage{amsmath}
\usepackage{amsfonts}
\usepackage{amssymb}
\usepackage{bm}
\usepackage{bbm}

\begin{document}

\begin{center}
{\large \bf Non-uniform convergence of two-photon decay rates
for excited atomic states}\\[2ex]
\normalsize
Ulrich D. Jentschura\\[1ex]
\scriptsize 
{\it Max--Planck--Institut f\"ur Kernphysik,
Saupfercheckweg 1, 69117 Heidelberg, Germany}\\[4ex]
\normalsize
\begin{minipage}{16.0cm}
{\underline{Abstract}}
Two-photon decay rates in simple atoms such as hydrogenlike systems 
represent rather interesting fundamental problems in atomic physics.
The sum of the energies of the two emitted photons has to fulfill an 
energy conservation condition, the decay takes place via 
intermediate virtual states, and the total decay rate is 
obtained after an integration over the energy of one of the emitted 
photons. Here, we investigate cases with 
a virtual state having an energy intermediate
between the initial and the final state of the decay process, 
and we show that due to non-uniform convergence,
only a careful treatment of the singularities infinitesimally
displaced from the photon integration contour leads to consistent 
and convergent results. 

\end{minipage}
\end{center}

Two-photon decay in atomic systems continues to 
be of both theoretical and 
experimental interest today 
(see, e.g., Ref.~\cite{DuEtAl1997}). 
Here, we shall investigate a mathematical
subtlety of the problem, which ultimately reveals that correct
results for two-photon decay widths of highly excited states
depend on a careful analysis of the singularities close to the 
photon integration contour.
In general, the decay width of a bound system
may be understood naturally as 
the imaginary part of the self energy~\cite{BaSu1978}.
This because the fundamental time evolution of a 
Schr\"{o}dinger eigenstate of energy $E_i$, which 
reads $\exp(-{\rm i} E_i t)$, must be modified to 
$\exp[-{\rm i}\, (E_i + {\rm Re} \, \Delta E_i) \, t - 
\textstyle{\frac12} \, \Gamma \, t]$ once
a perturbation leads to both a real and an imaginary
energy shift according to 
\begin{equation}
E_i \to E_i + \Delta E_i\,, \qquad
\Delta E_i = {\rm Re} \, \Delta E_i - {\rm i} \, \frac{\Gamma}{2}\,.
\end{equation}
It is known~\cite{Je2004rad} that the imaginary part of the two-photon 
self-energy shift ${\rm Im} \Delta E_i = - \textstyle{\frac12} \,
\Gamma$ gives rise to the two-photon decay width.
% (and to further radiative corrections 
% to the one-photon decay, which can be ignored in the context of
% the current investigation). 
Important steps toward a full clarification
of the two-photon processes involving excited states have been
accomplished in Refs.~\cite{Fl1984,CrTaSaCh1986,FlScMi1988}. Here, it is 
our intention to clarify the role of intermediate, virtual
states, whose energy lies between the energy $E_i$ of the initial 
state and the energy $E_f$ of the final state of the decay process.
The concept developed in~\cite{BaSu1978} guides us in our 
investigation.

We start from the nonrelativistic two-loop self 
energy~\cite{Pa2001,Je2004rad} in natural units
($\hbar = c = \epsilon_0 = 1$), with $m$ denoting the electron mass and 
$\alpha$ the fine-structure constant,
\begin{equation}
\label{NRQED}
\Delta E_i = \lim_{\epsilon \to 0}
\left( \frac{2 \alpha}{3 \pi m^2} \right)^2
\int_0^{\Lambda_1} d\omega_1 \, \omega_1 
\int_0^{\Lambda_2} d\omega_2 \, \omega_2 \, 
f_\epsilon(\omega_1, \omega_2)\,.
\end{equation}
Here, $\Lambda_1$ and $\Lambda_2$ are ultraviolet cutoff parameters,
and in $f_\epsilon(\omega_1, \omega_2)$ we carefully keep 
track of all infinitesimal imaginary parts,
\scriptsize
\begin{align}
f_\epsilon(\omega_1, \omega_2) =&
\left< \phi_i \left| 
p^j \, \frac{1}{E - H - \omega_1 + {\rm i}\epsilon} \, p^k \, 
\frac{1}{E - H - \omega_1 - \omega_2 + {\rm i}\epsilon} \, p^j \, 
\frac{1}{E - H - \omega_2 + {\rm i}\epsilon} \, p^k  \right| \phi_i \right> 
\nonumber\\[1ex]
& + \frac{1}{2} \,
\left< \phi_i \left| 
p^j \, \frac{1}{E - H - \omega_1 + {\rm i}\epsilon} \, p^k \,
\frac{1}{E - H - \omega_1 - \omega_2 + {\rm i}\epsilon} \, p^k \, 
\frac{1}{E - H - \omega_1 + {\rm i}\epsilon} \, p^j  \right| \phi_i \right> 
\nonumber\\[1ex] 
& + \frac{1}{2} \,
\left< \phi_i \left| 
p^j \, \frac{1}{E - H - \omega_2 + {\rm i}\epsilon} \, p^k \, 
\frac{1}{E - H - \omega_1 - \omega_2 + {\rm i}\epsilon} \, p^k \, 
\frac{1}{E - H - \omega_2 + {\rm i}\epsilon} \, p^j  \right| \phi_i \right>
\nonumber\\[1ex]  
& + 
\left< \phi_i \left| 
p^j \, \frac{1}{E - H - \omega_1 + {\rm i}\epsilon} \, p^j \, 
\left( \frac{1}{E - H} \right)' \, p^k \, 
\frac{1}{E - H - \omega_2 + {\rm i}\epsilon} \, p^k  \right| \phi_i \right> 
\nonumber\\[1ex]  
& - \frac{1}{2} \,
\left< \phi_i \left| 
p^j \, \frac{1}{E - H - \omega_1 + {\rm i}\epsilon} \, p^j  
\right| \phi_i \right> \,
\left< \phi_i \left| 
p^k \, \left( \frac{1}{E - H - \omega_2 + {\rm i}\epsilon} \right)^2 \, 
p^k  \right| \phi_i \right>
\nonumber\\[1ex]  
& - \frac{1}{2} \,
\left< \phi_i \left| 
p^j \, \frac{1}{E - H - \omega_2 + {\rm i}\epsilon} \, 
p^j  \right| \phi_i \right> \,
\left< \phi_i \left| 
p^k \, \left( \frac{1}{E - H - \omega_1 + {\rm i}\epsilon} \right)^2 \, 
p^k  \right| \phi_i \right>
\nonumber\\[1ex]
& + m \,
\left< \phi_i \left| p^j \, \frac{1}{E - H - \omega_1} \, 
\frac{1}{E - H - \omega_2 + {\rm i}\epsilon} \, p^j  \right| \phi_i \right>
- \frac{m}{\omega_1 + \omega_2} \,
\left< \phi_i \left| 
p^j \, \frac{1}{E - H - \omega_2 + {\rm i}\epsilon} \, p^j  
\right| \phi_i \right>
\nonumber\\[1ex]
& - \frac{m}{\omega_1 + \omega_2} \,
\left< \phi_i \left| 
p^j \, \frac{1}{E - H - \omega_1 + {\rm i}\epsilon} \, p^j  
\right| \phi_i \right> \,.
\end{align}
\normalsize
Here, $E_i$ is the Schr\"{o}dinger energy of the reference 
state, which is qualified here as the initial
state of the two-photon decay process, 
and $H$ is the Schr\"{o}dinger Hamiltonian.
Sums over the Cartesian coordinates $j,k \in \{ 1,2,3 \}$ 
are implied throughout this communication (summation convention).
One may ask why the infinitesimal imaginary parts in the 
propagator denominators have such a sign that $E_i$ effectively 
seems to 
acquire a positive imaginary part. That is not the case:
the reference state is assumed to be an asymptotic state in this 
formalism and does not have any imaginary part associated to it
at all. The infinitesimal imaginary parts are due to the 
virtual states (included in $H$), and these acquire an infinitesimal
negative imaginary part, as they should.

If the reference state is an excited state, then various singularities
are encountered along both the $\omega_1$ and $\omega_2$ 
integrations. As shown in Ref.~\cite{Je2004rad},
the two-photon decay rate can be obtained from the imaginary 
part of the two-loop self energy, upon consideration of 
those imaginary parts which are generated when a virtual 
state $| \phi_v \rangle$ with energy $E_v$ and the two emitted
photons meet at a resonance condition: $E - E_v = \omega_1 + \omega_2$.
At these points, expressions of the type $1/(E - H - \omega_1 - \omega_2)$
become singular. 

In order to allow for a consistent treatment
of the two-photon decay rate of excited states, it is necessary
to treat the energy shift (\ref{NRQED}) as a whole, to carefully keep
track of all ${\rm i} \epsilon$ terms, and to defer the 
distinction of imaginary and real parts to a later point.
Carrying out the integration over one of the photon energies
using the Dirac prescription
\begin{equation}
\frac{1}{a - \omega + {\rm i}\epsilon} = 
- {\rm i} \pi \, \delta(\omega - a) + ({\rm P}) \frac{1}{a - \omega} \,,
\end{equation}
where $({\rm P})$ denotes the principal value, we find that the 
two-photon decay rate corresponds to the expression
\begin{equation}
\label{decay}
\frac{\Gamma}{A} = \lim_{\epsilon \to 0} \, 
{\rm Re}
\int\limits_0^{E_i - E_f}
{\rm d}\omega \, \omega \, (E_i - E_f - \omega) \, 
\left\{ \left< \phi_f \left| p^j \, 
\frac{1}{E_i - H - \omega + {\rm i} \epsilon} \, 
p^j \right| \phi_i \right> +
\left< \phi_f \left| p^j \, 
\frac{1}{E_f - H + \omega + {\rm i} \epsilon} \,
p^j \right| \phi_i \right> \right\}^2 \,.
\end{equation}
where 
\begin{equation}
\label{defA}
A = \frac{4}{27}\,\frac{\alpha^2}{\pi}\,,
\end{equation}
In general, the expression (\ref{decay}) has both a 
real and an imaginary part (where it not for the enforced selection
of the real part implied by the ``Re'' in the cited equation). 
In that context, it is useful to observe 
that $\Gamma$ already 
manifests itself as the imaginary part of the energy shift (\ref{NRQED}).
The real part of $\Gamma$, in turn, gives the 
decay rate, and by consequence the ``imaginary part 
of $\Gamma$'' corresponds to a 
real part of the original energy shift (\ref{NRQED}), which 
is of the ``squared decay-rate'' type discussed 
in~\cite{JeEvKePa2002,Je2006}. The structure of the energy
shift associated with a resonance is quite intriguing in higher orders.

In a basis-set representation, Eq.~(\ref{decay}) reads
\begin{equation}
\label{decay_basis}
\frac{\Gamma}{A} = \lim_{\epsilon \to 0} \, 
{\rm Re}
\int\limits_0^{E_i - E_f}
{\rm d}\omega \, \omega \, (E_i - E_f - \omega) \, 
\sum_v 
\left\{ 
\frac{\left< \phi_f \left| p^j \right| \phi_v \right> 
\left< \phi_v \left| p^j \right| \phi_i \right> }
  {E_i - E_v  - \omega + {\rm i} \epsilon} +
\frac{\left< \phi_f \left| p^j \right| \phi_v \right> 
\left< \phi_v \left| p^j \right| \phi_i \right> }
{E_f - E_v + \omega + {\rm i} \epsilon} \right\}^2 \,,
\end{equation}
where the sum over $v$ contains all virtual states,
i.e.~over the entire bound and continuous spectrum.

The expression (\ref{decay}) now gives us a clear 
prescription how to handle the potentially problematic 
case of a virtual state $| \phi_v \rangle$ having an intermediate 
energy $E_v$ with $E_f > E_v > E_i$. An example is a virtual 
$|\phi_v \rangle = | 2{\rm P} \rangle$ state for a 
reference state $|\phi_i \rangle = | 3{\rm S} \rangle$
and a final state  $|\phi_f \rangle = | 1{\rm S} \rangle$. 
The treatment can be illustrated 
in a very clear manner by investigating the general
structure of the terms generated by the virtual states 
with intermediate energies $E_f > E_v > E_i$. 
We treat the square of the 
first term in curly brackets in Eq.~(\ref{decay_basis})
as an example. 
Indeed, after an appropriate scaling of the photon energy
integration variable, the expression 
takes the following form ($0 < a < 1$),
\begin{equation}
\label{model}
\lim_{\epsilon \to 0}  {\rm Re}
\int_0^1 {\rm d}\omega \, 
\left( \frac{1}{a - \omega + {\rm i}\epsilon} \right)^2 = 
\lim_{\epsilon \to 0} 
\int_0^1 {\rm d}\omega \, 
\frac{(a - \omega)^2 - \epsilon^2}{[(a - \omega)^2 + \epsilon^2]^2} =
\lim_{\epsilon \to 0} 
\left(\frac{a-1}{(a-1)^2 +\epsilon^2} - \frac{a}{a^2 + \epsilon^2} 
\right) = \frac{1}{a(a-1)} \,.
\end{equation}
This result holds strictly for $0 < a < 1$, but the limit is not
approached uniformly, i.e.~it would be forbidden to exchange the 
sequence of the limit $\epsilon \to 0$ with the integration over 
$\omega$. Similar phenomena are observed in the context of 
the renormalization of quantum electrodynamic processes, which 
necessitate the preservation of all relativistically covariant
regulators up to the very end of the calculation,
also due to non-uniform convergence.
Having Eqs.~(\ref{decay}) and (\ref{decay_basis}),
it is easy to perform actual numerical evaluations
of the two-photon decay rates for hydrogenlike systems.
Numerical results are given in Table~\ref{results}.

\begin{table}[thb]
\begin{center}
\begin{minipage}{13cm}
\begin{center}
\caption{\label{results} Two-photon decay rates obtained for
hydrogenlike systems by evaluation of Eq.~(\ref{decay}).
Various excited $n{\rm S}$ states are considered, both
as initial (excited) and as final states of the two-photon 
process. The results given here are for nuclear charge $Z=1$
and scale with $Z^6$ for hydrogenlike systems. Units are
inverse seconds. To obtain the decay rate in Hertz, one divides 
by a factor of $2\pi$.}
\begin{tabular}{c@{\hspace{0.8cm}}c@{\hspace{0.8cm}}c@{\hspace{0.8cm}}%
c@{\hspace{0.8cm}}c}
\hline
\hline
\rule[-2mm]{0mm}{6mm}
& 
$|\phi_i \rangle = | 2{\rm S} \rangle$ & 
$|\phi_i \rangle = | 3{\rm S} \rangle$ & 
$|\phi_i \rangle = | 4{\rm S} \rangle$ & 
$|\phi_i \rangle = | 5{\rm S} \rangle$ \\
\hline
\rule[-2mm]{0mm}{6mm}
$|\phi_f \rangle = | 1{\rm S} \rangle$ & 
  $8.229\,352$ & $2.082\,853$ & $0.698\,897$ & $0.287\,110$ \\
\rule[-2mm]{0mm}{6mm}
$|\phi_f \rangle = | 2{\rm S} \rangle$ & 
  $-$          & $0.064\,530$ & $0.016\,840$ & $0.001\,809$ \\
\rule[-2mm]{0mm}{6mm}
$|\phi_f \rangle = | 3{\rm S} \rangle$ & 
  $-$          & $-$          & $0.002\,925$ & $0.000\,704$ \\
\rule[-2mm]{0mm}{6mm}
$|\phi_f \rangle = | 4{\rm S} \rangle$ & 
  $-$          & $-$          & $-$          & $0.000\,297$ \\
\hline
\hline
\end{tabular}
\end{center}
\end{minipage}
\end{center}
\end{table}

By contrast, let us suppose
we had replaced the right-hand side of (\ref{decay}) 
by the expression
\begin{equation}
\int\limits_0^{E_i - E_f}
{\rm d}\omega \, \omega \, (E_i - E_f - \omega) \, 
\left| \left< \phi_f \left| p^j \, 
\frac{1}{E_i - H - \omega + {\rm i} \epsilon} \, 
p^j \right| \phi_i \right> +
\left< \phi_f \left| p^j \, 
\frac{1}{E_f - H + \omega + {\rm i} \epsilon} \,
p^j \right| \phi_i \right> \right|^2 \,,
\end{equation}
which has no physical meaning.
We have enforced a real valued integrand 
by the introduction of the complex modulus.
In that case, we would have obtained a divergent integral
in the limit $\epsilon \to 0$, because
\begin{equation}
{\rm Re} \int_0^1 {\rm d}\omega \, 
\left| \frac{1}{a - \omega + {\rm i}\epsilon} \right|^2 = 
\frac{\pi}{\epsilon} + \frac{1}{a(a-1)} + 
\frac13\, \left(\frac{1}{a^3} - \frac{1}{(a-1)^3}\right) \,
\epsilon^2 +
{\cal O}(\epsilon^4) \,.
\end{equation}
An infinite two-photon decay rate cannot be considered
physically sensible.

We can thus conclude that virtual intermediate 
states with energies that lie between the energy of the 
initial and final states of a two-photon decay process contribute
a finite correction to the two-photon decay rate,
although the integrand of Eq.~(\ref{model}), in the 
limit $\epsilon \to 0$, has a non-integrable singularity 
at $\omega = a$ of the form $1/(a-\omega)^2$. 
The convergence of the integral (\ref{model}) in the limit $\epsilon \to 0$
is not uniform, and our example shows that the concept
of non-uniform convergence is not merely a mathematical 
sophistication in the context of two-photon processes: 
it ensures that finite, physically sensible results are
obtained for the two-photon decay rates, which include the contribution
from all possible virtual states.
A generalization of the results obtained in this communication
to the relativistic and to the many-electron case
is straightforward. Finally, it should be remarked that 
at least in principle,
the two-photon decay rates can be measured experimentally 
although they are orders of magnitude smaller than 
one-photon rates for many of the processes listed in 
Table~\ref{results}. The point is that only in two-photon decay,
both photons can be detected in coincidence, and events can be 
selected by considering the center-of-mass energy.

The author acknowledges support from Deutsche Forschungsgemeinschaft
(Heisenberg program).

\end{document}